\documentclass[12pt]{iopart}

\usepackage{graphics}
\usepackage{bm}
\usepackage{color}

\def\PRA{{\it Phys.~Rev.~A} }

\def\JPB{{\it J.~Phys.~B} }
\def\PRL{{\it Phys.~Rev.~Lett.} }
\def\RMP{{\it Rev.~Mod.~Phys.} }
\def\JCP{{\it J.~Chem.~Phys.} }

\newcommand{\myscaleboxa}[1]{\scalebox{0.95}[0.95]{#1}}
\newcommand{\myscaleboxb}[1]{\scalebox{0.425}[0.425]{#1}}
\newcommand{\myscaleboxc}[1]{\scalebox{1.2}[1.2]{#1}}
\newcommand{\myscaleboxd}[1]{\scalebox{0.5}[0.5]{#1}}

\newcommand{\be}{\begin{equation}}
\newcommand{\bea}{\begin{eqnarray}}
\newcommand{\eea}{\end{eqnarray}}
\newcommand{\ee}{\end{equation}}

\begin{document}

\title{Uncovering multiple orbitals influence in high
harmonic generation from aligned N$_2$}

\author{Anh-Thu Le,$^1$ R.~R. Lucchese,$^2$ and C.~D. Lin$^1$}

\address{$^1$ Department of Physics, Cardwell Hall, Kansas
State University, Manhattan, KS 66506, USA}

\address{$^2$ Department of Chemistry, Texas A\&M University, College Station,
Texas 77843-3255, USA}

\date{\today}

\begin{abstract}
Recent measurements on high-order harmonic generation (HHG) from
N$_2$ aligned perpendicular to the driving laser polarization [B.~K.
McFarland {\it el al}, Science {\bf 322}, 1232 (2008)] have shown a
maximum at the rotational half-revival. This has been interpreted as
the signature of the HHG contribution from the molecular orbital
just below the highest occupied molecular orbital (HOMO). By using
the recently developed quantitative rescattering theory combined
with accurate photoionization transition dipoles, we show that the
maximum at the rotational half-revival is indeed associated with the
HOMO-1 contribution. Our results also show that the HOMO-1
contribution becomes increasingly more important near the HHG cutoff and
therefore depends on the laser intensity.
\end{abstract}

 \submitto{\JPB} \maketitle

High-order harmonic generation (HHG) has been extensively
investigated both experimentally and theoretically over the last two
decades \cite{krausz09}. Until very recently, the HHG has been understood as being due to
tunneling ionization of an electron from the highest occupied
molecular orbital (HOMO) and recombining back to the HOMO. The
contribution from lower molecular orbitals are routinely neglected.
That is not surprising since tunneling is a highly nonlinear
process, and therefore highly selective to the HOMO due to
energy considerations. In general, the neglect of the contribution
from lower molecular orbitals are not justified for systems
where the HOMO and HOMO-1 are nearly degenerate,
 in other words, when the energy gap between
the HOMO and lower molecular orbitals are much smaller than the
ionization potential from the HOMO. Furthermore, for some molecular
alignments, tunneling ionization from the HOMO is suppressed due to
symmetry of the wavefunction \cite{moadk}. Clearly, in that case the neglect of
lower molecular orbitals is questionable. These two favorable
conditions for observing a HOMO-1 contribution are present in N$_2$,
where the 1$\pi_u$ HOMO-1 has a binding energy of 16.93 eV, quite
close to the binding energy of the 3$\sigma_g$ HOMO (15.58 eV).

Early theoretical calculations based on the strong-field approximation (SFA) model
\cite{zhou05,madsen06} have shown that harmonic yields from aligned
N$_2$ are maximum if the molecules are aligned along the laser
polarization direction. These results are in good agreements with
the pump-probe delay time experimental data \cite{itatani05,kanai05}
as well as the recent more direct measurements \cite{mairesse08}. On
the other hand, the contribution from the HOMO-1 is expected to peak
near 90$^{\circ}$. Although the two molecular orbitals contribute to
different alignment regions in the total harmonic yields, it is
still a very challenging task to disentangle the HOMO-1 since its
contribution is expected to be relatively weak. This is in strong
contrast to the traditional single-photon photoionization where
electrons are generally ionized from many MOs with comparable strengths
(see, for example, \cite{thomann08}).

In a recent experiment McFarland {\it et al} \cite{mcfarland}
reported that they have successfully observed the contribution from
the HOMO-1 in aligned N$_2$. That has been achieved within the
pump-probe scheme with perpendicular pump-probe polarizations. For
low harmonic orders below ~H23, the harmonic signals behave
similarly to {\em inverted} $<\cos^2\theta>$, i.e., inverse of the degree of
molecular alignment or $1-<\cos^2\theta>$. For higher harmonics, McFarland {\it et al}
observed a maximum at the rotational half-revival, where {\em inverted}
$<\cos^2\theta>$ is minimum. Furthermore, the maximum at
the half-revival is found to be quite pronounced in the HHG cutoff
region, the location of which depends on the intensity of the driving laser.

The goal of this paper is to show theoretically that the main
features observed by McFarland {\it et al} \cite{mcfarland} are
indeed the signature of the HOMO-1 contribution. To support our
claim, we have carried out calculations by using the recently
developed quantitative rescattering theory (QRS) \cite{atle09-long}.
The photoionization transition dipole and its phase are obtained
from state-of-the-art molecular photoionization calculations
\cite{lucchese95}. The QRS theory is based on the rescattering
picture and it has been shown to give accurate results comparable
with that from the time-dependent Schr\"odinger equation (TDSE) for
rare-gas atoms \cite{toru08,atle08} and the molecular ion H$_2^+$
\cite{H2+}. The QRS has also been shown to be able to reproduce most
of the available experiments on aligned molecules CO$_2$, O$_2$ and
N$_2$ \cite{atle09,atle09-long}. Other applications of the QRS
include calculations of high-energy above-threshold ionization
momentum and energy spectra (see, for example, \cite{chen09} and
references therein) and nonsequential double ionization
\cite{sam-nsdi}. Analytical derivations for the QRS have been
reported quite recently in
Refs.~\cite{frolov09a,frolov09b,becker09}. In this paper we extend
the QRS theory \cite{atle09-long} to the multi-channel case, where
the contribution from each channel to the total HHG induced dipole
are added coherently. We assume that the ion cores are frozen during
the time interval between ionization and photo-recombination.

\begin{figure}
\centering \mbox{\rotatebox{0}{\myscaleboxa{
\includegraphics{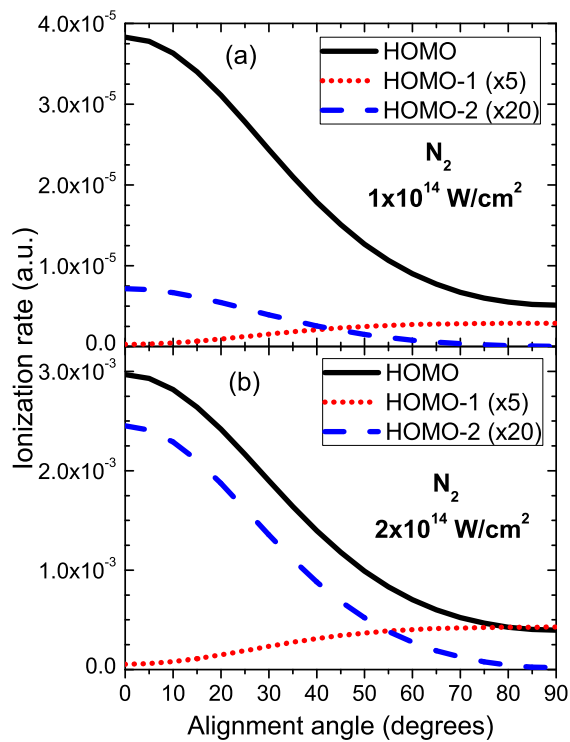}}}}

\caption{Ionization rates from HOMO, HOMO-1, and HOMO-2 at laser
intensities of $1\times 10^{14}$ (a) and $2\times 10^{14}$ W/cm$^2$
(b). The data from HOMO-1 and HOMO-2 have been multiplied by a
factor of 5 and 20, respectively. Calculations are carried out
within the MO-ADK theory.} \label{fig1}
\end{figure}

First we note that the HOMO ($3\sigma_g$), HOMO-1 ($1\pi_u$) and
HOMO-2 ($2\sigma_u$) have binding energies of 15.58, 16.93 and
18.73 eV, respectively. In order to have an idea about the magnitude
of the relative contributions from the HOMO, HOMO-1 and HOMO-2 in the
returning electron wave packet we compare the ionization rates from
these molecular orbitals. These comparisons are shown in Fig.~1 for
two different laser intensities of $1\times 10^{14}$ and $2\times
10^{14}$ W/cm$^2$. The calculations were performed within the
molecular tunnelling (MO-ADK) theory \cite{moadk}. The MO-ADK $C_l$
coefficients were calculated from asymptotic wavefunctions of each
orbital, which in turn were obtained from the {\it GAUSSIAN} code
\cite{gaussian}. Clearly the ionization rates depend strongly on the
alignment and the alignment dependent rates are different for
different symmetries of the orbitals. This fact has been known
before for HOMOs both theoretically \cite{zhao03,faisal00} and
experimentally \cite{NRC03,cocke04}. For the HOMO-1 ($1\pi_u$) the
ionization rate peaks near $90^{\circ}$. Within this range of laser
intensity, the rate from the HOMO-1 is a factor of 5 smaller than that
of the HOMO, even at $90^{\circ}$. For the HOMO-2 ($2\sigma_u$) the
ionization rate peaks near $0^{\circ}$, similar to that of the HOMO.
However its magnitude is significantly smaller than that of the HOMO
(approximately by a factor of 20 for this range of laser intensity).
We note that the relative contributions from the lower orbitals
become increasingly more important as the laser intensity increases.
We also carried out calculations using the SFA and found similar
angular dependence for ionization rates from these three MOs. The
relative strengths are also in good agreements with the MO-ADK,
although the rate for the HOMO-2 from the SFA seems to be enhanced by a
factor of 5, (i.e., still about a factor of 10 weaker than the
HOMO). We note that our results do not agree with the recent results
from the time-dependent density-functional theory calculations
\cite{telnov09}, which show a peak in ionization rate from the HOMO-1 at
an intermediate angle.

\begin{figure}
\centering \mbox{\rotatebox{0}{\myscaleboxb{
\includegraphics{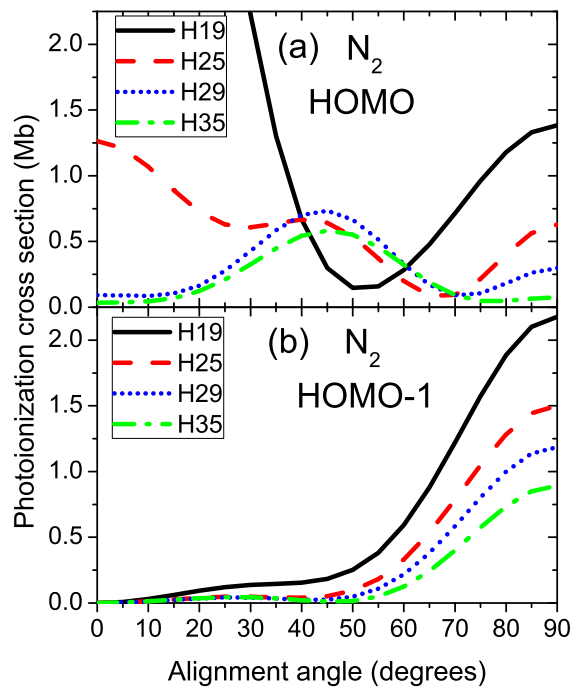}}}}

\caption{Differential photoionization cross sections vs alignment
angle from HOMO (a) and HOMO-1 (b) for some selected energies
(expressed in units of harmonic orders for an 800-nm laser).}
\label{fig2}
\end{figure}

The above analysis shows that the contribution to the returning wave
packet from the HOMO-1 is about a factor of 5 smaller than that of the
HOMO even near its peak at $90^{\circ}$. Can the HOMO-1 contribution
to the harmonic generation be comparable with that of the HOMO? To
answer this question we show in Fig.~2 a comparison of the
differential photoionization cross sections from the HOMO and
HOMO-1. (In this paper we limit ourselves to the harmonics with
polarization parallel to that of the driving laser. Therefore the
relevant differential cross sections are for the electron emitted along
the laser polarization direction \cite{atle09-long}.)  For
convenience we express photon energy in units of the photon energy
of the 800-nm laser (1.55 eV). First we note that the cross sections
from the HOMO vary strongly from one energy to the next.
Nevertheless, one general feature can be seen is that the cross
sections at large angles near $90^{\circ}$ are quite small, say,
about 0.25 Mb for H29, and decreases quickly with energy. On the other
hand the photoionization cross sections from the HOMO-1 all have a
dominant peak at $90^{\circ}$, which reaches 1.25 Mb for H29.
Recall that according
to the QRS theory, HHG yield is proportional to a product of the
returning wave packet and the differential photoionization cross
section. The results shown in Figs.~1 and 2 therefore indicate that
the HOMO and HOMO-1 contributions could be comparable near
$90^{\circ}$.

\begin{figure}
\centering \mbox{\rotatebox{0}{\myscaleboxc{
\includegraphics{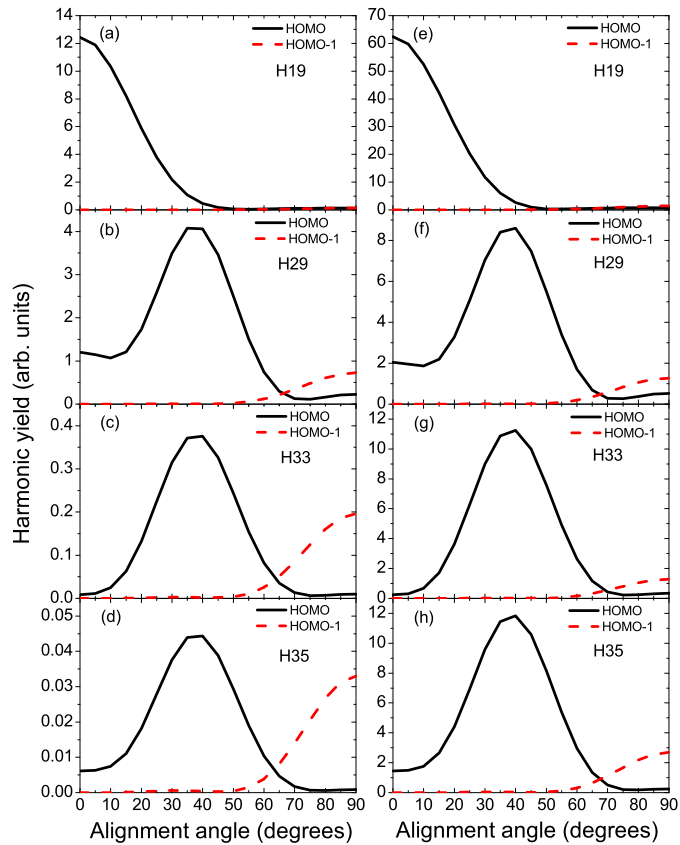}}}}

\caption{Contributions to HHG yields from HOMO and HOMO-1 for some
selected harmonics with the laser intensity of $1.5\times 10^{14}$
W/cm$^2$ (left column) and $2\times 10^{14}$ W/cm$^2$ (right
column). Molecular axis is assumed to be fixed (no averaging over
alignment distribution was carried out).} \label{fig3}
\end{figure}

Having established qualitatively that the contribution from the HOMO-1
cannot be neglected for large angles, we now analyze the HHG yields
from the actual QRS calculations. In Fig. 3(a-d) (left column) we
show the HHG yields from the HOMO and HOMO-1 for H19, H29, H33 and H35.
The 800-nm laser pulse is of 30fs duration (FWHM) and intensity of
$1.5\times 10^{14}$ W/cm$^2$. Clearly, the HOMO contribution
dominates for alignment angles smaller than $45^{\circ}$ for all the
harmonics. For large angles the HOMO-1 becomes comparable with the
HOMO already near H25 and dominates for the higher harmonics,
especially beyond the cutoff at H29. Similar pattern repeats at a
higher laser intensity of $2\times 10^{14}$ W/cm$^2$, shown in
Fig.~3(e-h) (right column). However the HOMO-1 is only comparable
with the HOMO at large angles for H29 and it starts to dominate only at
higher harmonics. The enhanced contribution from the HOMO-1 near the
cutoff can be understood as the consequence of the delay in the
harmonic cutoff for the HOMO-1, since the ionization potential from the
HOMO-1 is about 1.5 eV greater than that from the HOMO. As the
cutoff moves to near H35 for the laser intensity of $2\times
10^{14}$ W/cm$^2$, the HOMO-1 contribution dominates at much higher
harmonic orders as compared to the lower intensity case shown in the left
column. This fact has been noticed earlier by McFarland {\it et al}
\cite{mcfarland}.

\begin{figure}
\centering \mbox{\rotatebox{0}{\myscaleboxd{
\includegraphics{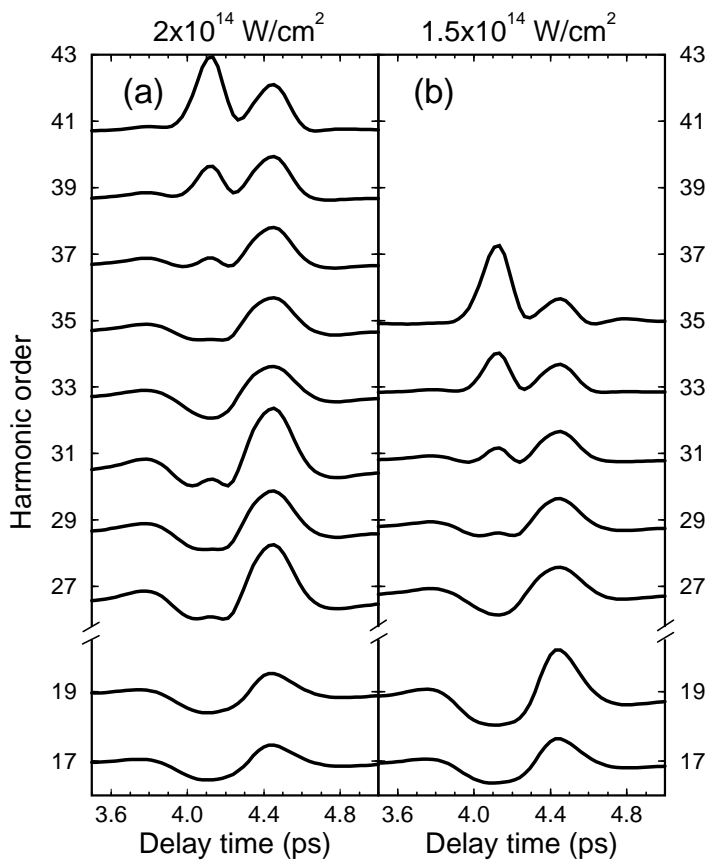}}}}
\caption{Harmonic signals as functions of delay time near
half-revival. The laser intensity is $2\times 10^{14}$ W/cm$^2$ (a)
and $1.5\times 10^{14}$ W/cm$^2$ (b). A (pump) laser pulse of 90 fs
duration (FWHM), intensity of $3\times 10^{13}$ W/cm$^2$ is used to
align the molecules.} \label{fig4}
\end{figure}

The main results of this paper are presented in Fig.~4, where we
show the harmonic signals as functions of delay time between the
pump and probe laser pulses at two intensities of $1.5\times
10^{14}$ and $2\times 10^{14}$ W/cm$^2$. The signals have been
normalized to the signals from the isotropic distribution. These
results can be compared directly with the measured HHG signals by
McFarland {\it et al} \cite{mcfarland}. Similarly to the experiments
by McFarland {\it et al} \cite{mcfarland}, we take the pump laser
polarization to be perpendicular to the probe laser polarization.
This is used in order to facilitate the observation of the harmonic
signals at the large angles where the HOMO-1 is dominant for high
harmonic orders. Theoretically the harmonic signals are obtained by
a coherent convolution of the HHG yields from the QRS calculations
with the partial alignment distribution. The time-dependent
molecular alignment distribution is calculated by solving the TDSE
for N$_2$ molecules in the pump (alignment) laser field within the
rotor model \cite{seideman}. The pump (alignment) laser has
a pulse length of 90 fs
(FWHM), intensity of $3\times 10^{13}$ W/cm$^2$ and 800-nm
wavelength. We assume the Boltzmann distribution for the rotational
levels at the initial time and the rotational temperature is taken
to be 40 K. The above parameters are chosen to closely match the
experimental conditions of McFarland {\it et al} \cite{mcfarland}.

At lower laser intensity of $1.5\times 10^{14}$ W/cm$^2$, as can be
seen from Fig.~4(b), the lower harmonic (as represented by H17 and
H19) has a minimum at the half-revival near 4.1 ps. In other words, the HHG
signal behaves as inverted $<\cos^2\theta>$ (not shown), which
measures the degree of molecular alignment. For H29 and higher
harmonics a peak superimposed on the minimum can be seen. This is
clear evidence for the increasing importance of the contribution
from the HOMO-1 at large angles, shown in Fig.~3(a). Recall that at the
half-revival near 4.1 ps, the molecules are maximally aligned along
the pump polarization direction, which is perpendicular to probe
polarization. If we artificially remove the HOMO-1 contribution, all
the harmonics behave similarly to H17, i.e., as inverted
$<\cos^2\theta>$. On the other hand, the contribution from the
HOMO-2 are found to be negligibly small and its inclusion does not
affect our results. Similar behavior is seen at higher laser
intensity of $2\times 10^{14}$ W/cm$^2$ shown in Fig.~4(a). However
the peak at 4.1 ps starts to show up only at H31 and more
systematically after H35. This is due to the fact that the harmonic
cutoff is shifted to H35 at this high intensity. It is clear that
our QRS theoretical results for a single-molecule response already
reproduce quite well the general behavior reported by McFarland {\it
et al} \cite{mcfarland}. We also performed calculations using the
standard SFA, with and without including the multiple orbitals, and
found that the SFA does not reproduce experimental results. For a
more complete theory, one certainly needs to include the macroscopic
propagation. As has been reported quite recently by Sickmiller and
Jones \cite{jones09}, the phase-matching condition can have a
dramatic effect on the HHG signal in the waveguide. The effect of
propagation is expected to be much less significant in the gas jet
experiment by McFarland {\it et al} \cite{mcfarland} than in
Ref.~\cite{jones09}. At present the macroscopic propagation of the
HHG in molecular gas media is still largely unexplored with only one
attempt for H$_2^+$ \cite{bandrauk07} having been reported so far.

In conclusion, by using the recently developed quantitative
rescattering theory (QRS) combined with accurate photoionization
transition dipoles we have confirmed theoretically that the peaks
superimposed on the minimum near rotational half-revival observed
experimentally in N$_2$ by McFarland {\it et al} \cite{mcfarland}
can be attributed to the contribution from the HOMO-1. Our results
also show that the contribution from the HOMO-1 becomes more
important in the HHG cutoff region and therefore depends on the
laser intensity. In this paper we have limited ourselves to the
parallel polarization component for the emitted harmonics, as only
the case of orthogonally polarized pump and probe pulses was
considered. A recent experiment by Zhou {\it et al} \cite{jila09}
and an earlier experiment by Levesque {\it et al} \cite{levesque07},
in which the angle between pump and probe polarizations were varied,
showed that the ratio of the perpendicular to parallel component
intensities could be as high as $\sim 0.2$. However this ratio
decreases significantly for large angles between the pump and probe
polarizations and goes to zero at $90^{\circ}$ due to the symmetry.
The importance of this effect is currently under investigation.
Finally we mention the recent experimental and theoretical results
by Smirnova {\it et al} \cite{smirnova09}, which presented evidence
for the multiple orbitals effect in CO$_2$.

We thank M. G{\"u}hr, B. McFarland, J. Farrell, and P. Bucksbaum for
communicating their experimental data to us and for valuable
discussions. This work was supported in part by the Chemical
Sciences, Geosciences and Biosciences Division, Office of Basic
Energy Sciences, Office of Science, U. S. Department of Energy.

\end{document}